\documentclass[]{aa}
\usepackage{graphicx,amssymb,lscape,setspace,xspace}
\begin{document}

%\thesaurus{11          % Galaxies
%(           11.06.2;   % galaxies: fundamental parameters
%            11.09.5;   % galaxies; irregular
%            13.19.1)} % radio lines: galaxies

\title{Ca\,{\sc ii} K observations of QSOs in the line-of-sight to 
the Magellanic Bridge 
       \thanks{Based on observations made with {\sc emmi} on the New Technology Telescope, La Silla, Chile, 
               programme ID=074.A-0038(A).}	   
       }

\author{J. V. Smoker
        \inst{1,2}
        \and
        F. P. Keenan
        \inst{1}
        \and
        H. M. A. Thompson
        \inst{1}
        \and
        C. Br\"uns
        \inst{3}
        \and
        E. Muller
        \inst{4}
        \and
        N. Lehner
        \inst{5}
        \and
        \\
        J.-K Lee
        \inst{1}
        \and
        I. Hunter
        \inst{1}
        }

\institute{APS division,
           Dept. of Pure and Applied Physics,
           Queen's University,
           University Road,
           Belfast, BT7 1NN,
           U.K. 
\and
           European Southern Observatory,
           Alonso de Cordova 3107,
           Casilla 19001,
           Vitacura,
           Santiago 19, Chile
\and
           Radioastronomisches Institut, 
           Universit\"at Bonn, Auf dem H\"ugel 71, 
           53121 Bonn, 
           Germany
\and
           Australia Telescope National Facility, 
	   CSIRO, PO Box 76, Epping N.S.W,
           1710, Australia
\and
           Department of Astronomy, 
           University of Wisconsin, 
           475 North Charter Street, 
           Madison, WI 53706,
           U.S.A.
%\and
%           Department of Physics and Astronomy, 
%           University of Wyoming, 
%           Laramie, 
%           Wyoming,
%           U.S.A.
   }
\offprints{j.smoker@qub.ac.uk}
\date{Received / Accepted}

\abstract{We describe medium-resolution spectroscopic observations taken with the ESO 
Multi-Mode Instrument ({\sc emmi}) in the Ca\,{\sc ii} K line 
($\lambda_{\rm air}$=3933.661\AA) towards 7 QSOs located in the line-of-sight to the Magellanic 
Bridge. At a spectral resolution $R$ = $\lambda$/$\Delta\lambda$ = 6,000, five of the 
sightlines have a signal-to-noise (S/N) ratio of $\sim$ 20 or higher. Definite Ca 
absorption due to Bridge material is detected towards 3 objects, with probable detection 
towards two other sightlines. Gas-phase Ca\,{\sc ii} K Bridge and Milky Way 
abundances or lower limits for the all sightlines are estimated by the use of Parkes 21-cm H\,{\sc i} 
emission line data. These data only have a spatial resolution of 14 arcminutes compared 
with the optical observations which have milli-arcsecond resolution. With this caveat, 
for the three objects with sound Ca\,{\sc ii} K detections, we find that the ionic abundance 
of Ca\,{\sc ii} K relative to H\,{\sc i}, $A$=log($N$(Ca\,K)/$N$(H\,{\sc i})) for low-velocity 
Galactic gas ranges from --8.3 to --8.8 dex, with H\,{\sc i} column densities varying from 
3$-$6$\times$10$^{20}$ cm$^{-2}$. For Magellanic Bridge gas, the values of $A$ are 
$\sim$ 0.5 dex higher, ranging from $\sim$ --7.8 to --8.2 dex, with 
$N$(H\,{\sc i})=1--5$\times$10$^{20}$ cm$^{-2}$. Higher values of $A$ correspond to lower 
values of $N$(H\,{\sc i}), although numbers are small. For the sightline towards B\,0251--675, 
the Bridge gas has two different velocities, and in only one of these is Ca\,{\sc ii} 
tentatively detected, perhaps indicating gas of a different origin or present-day 
characteristics (such as dust content), although this conclusion is uncertain and there 
is the possibility that one of the components could be related to the Magellanic Stream. 
Higher signal-to-noise Ca\,{\sc ii} K data and higher resolution H\,{\sc i} data are required to determine 
whether $A$ changes with $N$(H\,{\sc i}) over the Bridge and if the implied difference in 
the metalicity of the two Bridge components towards B\,0251-675 is real. 
}

\titlerunning{CaK observations of the Magellanic Bridge}
\maketitle

\section{Introduction}

The Magellanic Bridge is a filament of H\,{\sc i}, joining the Large and Small 
Magellanic Clouds (LMC, SMC), first detected and characterized by Hindman 
et al. (1961). This region has 
many interesting features, 
in particular that, although the gas density (generally $N$(H\,{\sc i}) 
$\sim$ 10$^{20-21}$ cm$^{-2}$; McGee \& Newton 1986) and metallicity  
(0.08 solar; Rolleston et al. 1999) are relatively low, it still contains a number of 
young B-type stars whose evolutionary ages and 
velocities make it likely that they were born in the Bridge 
(Hambly et al. 1994; Rolleston et. al 1999), perhaps in 
CO-containing clouds discovered 
by Muller et al. (2003a). The Bridge may have 
been formed by a tidal interaction between the LMC and SMC some 200 Myr ago 
(e.g. Sawa et al. 1999).

\begin{figure*}
\begin{center}
\includegraphics[]{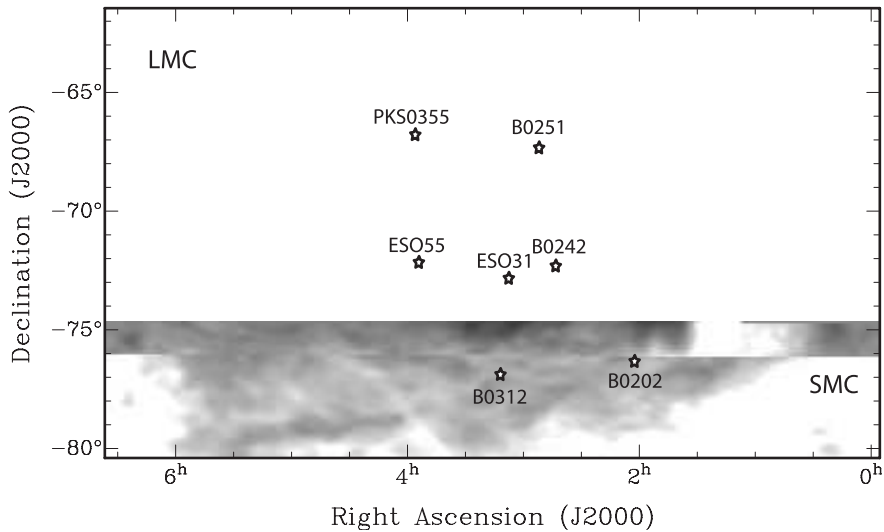}
\caption{Location of the QSO sample superimposed on the H\,{\sc i} column density 
distribution from Br\"uns et al. (2005). The Ca\,{\sc ii} spectra of ESO 31-8 and PKS 0355-66 
have low S/N ratios.
}
\label{samplefig}
\end{center}
\end{figure*}

Previous observations of the Bridge have concentrated on its stellar content and 
stellar associations (e.g. Irwin et al. 1990, Bica \& Schmitt 1995, Graham et al. 2001), 
stellar abundances (Rolleston et al. 1999), H\,{\sc i} content and dynamics 
(McGee \& Newton 1986, Putman et al. 1998, Kobulnicky \& Dickey 1999, 
Muller et al. 2003b) and CO content (Smoker et al. 2000, Muller et al. 2003a). 
However, to date there have only been two UV/optical interstellar absorption-line studies of the Bridge 
with the aim of determining the properties of the gas within it (Lehner et al. 2001, Lehner 2002). 
These studies used early-type stars located 
in the Bridge to probe the material in the lines-of-sight. 
The main findings are that a wide variety of ionisation 
stages are present (ranging from H$_{2}$ to Si\,{\sc iv} and C\,{\sc iv}), with 
the depletion patterns relative to S of Si, Fe and Ni being very similar to that
found for the Galactic 
halo. Although important, these studies suffer from the disadvantages  
that: (1) there are few sufficiently bright and blue stars within the Bridge 
for which optical or 
UV interstellar absorption measurements are possible, and (2) 
by their nature, it is unclear where
the stars lie within the depth of the Bridge, so comparing results 
to H\,{\sc i} emission-line 
observations which measure the integrated sightline 
is problematic. For these reasons, observing true 
background sources such as QSOs 
to determine the gas properties appears a worthwhile adventure. 

The current paper aims to improve our 
knowledge of Bridge gas by performing Ca\,{\sc ii} K spectroscopy 
towards a number of background QSOs. It is a pilot study, part of whose purpose is to 
provide objects for future higher spectral resolution studies, as {\em a priori} 
one may expect that the Bridge gas will be clumpy, with some sightlines showing little or no 
optical or UV interstellar absorption. The clumping exists at all measured scales, 
and is what would be expected for a fluid which has energy dumped in at large scales,
for example gravitationally (Muller et al. 2004). 

The paper is organized as follows: 
Section \ref{sample} describes the sample selection. 
The NTT Ca\,{\sc ii} K and Parkes-ATCA H\,{\sc i} 
observations and data reduction are elucidated in Sect. \ref{observations}, 
while Sect. \ref{analysis} 
describes the analysis of the observed Ca\,{\sc ii} K profiles and 
presents these spectra as well as 
the corresponding H\,{\sc i} observations. 
In Sect. \ref{discussion} we discuss the derived parameters as 
a function of position within the 
Bridge. Finally in Sect. \ref{conclusions} we give the main conclusions 
and avenues for future research. 

\section{The sample}
\label{sample}
The sample of QSOs observed is shown in Table \ref{sampletable}. Objects were 
selected from 
Hewitt \& Burbidge (1993) and Geha et al. (2003) and are a mix of 
Quasars and Seyfert galaxies. 
They were chosen so that they were observable with the NTT in a reasonable time, 
and also so 
that they spanned as large an angular area
of the Magellanic Bridge as possible. The location of the 
selected objects is shown in Fig. \ref{samplefig} which shows the 
sample objects superimposed on an H\,{\sc i} column density map of the Magellanic Clouds 
from Br\"uns et al. (2005).

\section{Observations and data reduction}
\label{observations}

\subsection{NTT {\sc emmi} Ca\,{\sc ii} K observations}

The optical observations described in this paper were taken 
during three nights in 2004 October 19-21 using the
ESO Multi-Mode Instrument ({\sc emmi}) on the 3.6-m 
New Technology Telescope (NTT) located on La Silla, Chile. 
Conditions for the first two nights were clear, with observations on 
night three when B\,0251--675 and 
PKS \,0355--66 were observed being affected by cloud. 
The seeing varied from 0.5--1.6 arcsec. The 
blue arm of {\sc emmi} 
was used with a pixel scale of 0.37 arcseconds 
per pixel with the holographic grating \#11 and 
slow readout in order to minimise readout noise (RON). 
The resulting RON was 5.0 electrons with a gain of 1.43. 
In combination with a 1.5 arcsec slit, 
a spectral resolution $R$ (=$\lambda$/$\Delta\lambda$) of 6,000 
or 50 km\,s$^{-1}$ was obtained, centred 
on 3933.6\AA \, and having a wavelength coverage of $\sim$150\AA. 
This resolution enabled us to distinguish 
between Galactic and Magellanic Bridge absorption, the latter being
offset by $\sim$ +200 km\,s$^{-1}$. 
ThAr arc-line spectra were observed 
with each target to minimise uncertainties in the wavelength scale 
caused by shifts in the instrument due to rotation 
as the objects were tracked. With the exception of 
PKS\,0355--66, at least 3 exposures per target were 
taken in order to make easier the task of 
cosmic-ray removal. Standard reduction procedures 
within  {\sc iraf}{\it \footnote{ {\sc iraf}
is distributed by the National Optical Astronomy Observatories, U.S.A.}} 
were used to produce the 1-D spectra 
from the 2-D images. These included bias subtraction and flatfielding using 
a pixel map created from 
dome flatfields taken during the daytime, 
and wavelength calibration using the attached ThAr arcs. 
The stability of the spectrograph was tested by checking the position of 
an arc line over ten images 
taken over three nights, and the maximum velocity shift found was 
10 km\,s$^{-1}$. However this is an upper limit in  
the wavelength calibration error, 
as the attached arcs were taken at the same rotator angle as the 
observations, minimising shifts caused by flexure of the instrument. 
Extraction of the reduced 2-D 
images was performed using 
the {\sc doslit} package. Of the seven objects observed, 
five had a signal-to-noise ratio in the continuum 
of $\sim$ 20 or above, with the remaining two objects having 
poorer quality spectra. After initial 
reduction and extraction using {\sc iraf}, 
the co-added Ca\,{\sc ii} K spectra were imported into 
{\sc dipso} (Howarth et al. 1996) and converted to the kinematical Local Standard of 
Rest (LSR) using corrections generated by the program {\sc rv} (Wallace \& Clayton 1996). 
The continuum was removed by fitting low-order baselines to the extracted spectra, from which we 
also estimated the signal to noise ratios, which range from 6--31 in the region of Ca\,{\sc ii} K. 
We note that the continua in the extracted spectra were relatively smooth and well-behaved with 
for example no obvious jumps. 

% 0202-765   l,b=297.55, -40.05
% 0242-7229  l,b=291.81, -42.35
% 0251-675   l,b=286.78, -45.87
% ESO31-8    l,b=290.33, -40.79
% 0312-770   l,b=293.44, -37.55
% ESO 55-2   l,b=286.61, -38.67
% PKS0355-66 l,b=280.63, -41.49

\begin{table}
\small
\caption{The Sample. Coordinates, redshifts and magnitudes are from {\sc simbad}. The Obs. time column refers to the 
{\sc emmi} shutter-open time in hours.}
\label{sampletable}
\begin{center}
\begin{tabular}{lrrrr}
\hline
QSO                      &  RA hms                                        & $z$            & $m_{B}$ & Obs. time   \\
(Type)                   &  Dec. $^{\circ}$ $^{\prime}$ $^{\prime\prime}$ &                & $m_{V}$ & (hours)     \\
                         &  (J2000)                                       &                & (mag.)  &             \\
\hline
                         &                                                &                &         &             \\ 
B\,0202--765               &    02 \,02 \,13.7                              &  0.39          &   16.83 &  3.75       \\
\hspace*{0.2cm}(Sy1)     &  --76 \,20 \,03                                &                &   16.77 &             \\
B\,0242--7229              &    02 \,43 \,09.6                              &  0.12          &      -- &  2.75       \\
\hspace*{0,2cm}(Sy1)     &  --72 \,16 \,48.9                              &                &   15.9  &             \\
B\,0251--675               &    02 \,51 \,55.8                              &  1.00          &      -- &  7.00       \\
\hspace*{0.2cm}(Quasar)  &  --67 \,18 \,00                                &                &   17.5  &             \\
ESO\,31-8                  &    03 \,07 \,34.9                              &  0.03          &   15.48 &  1.75       \\
\hspace*{0,2cm}(Sy1)     &  --72 \,50 \,04                                &                &     --  &             \\
B\,0312--770               &    03 \,11 \,55.2                              &  0.22          &     --  &  2.25       \\
\hspace*{0.2cm}(Sy1)     &  --76 \,51 \,51                                &                &   15.9  &             \\
ESO\,55--2               &    03 \,54 \,02.0                              &  0.05          &   15.67  &  3.75       \\ 
\hspace*{0,2cm}(Sy2)     &  --72 \,08 \,04                                &                &     --  &             \\
PKS\,0355--66             &    03 \,55 \,48.0                              &  --            &   17.3  &  0.92       \\
\hspace*{0.2cm}(Quasar)  &  --66 45 33.4                                  &                &     --  &             \\
\hline
\end{tabular}
\end{center}
\normalsize
\end{table}

%
% Results from RV
%
% RV results;  B 0202-765 To_Helicentric add -10.0 km/s. To_LSR_Kinematic add -21 km/s
%              B 0242-7229               add  -8.3 km/s. To_LSR_Kinematic add -21 km/s
%              B 0251-675                add  -7.3 km/s. To_LSR_Kinematic add -21 km/s
%              ESO 31- 8                add  -7.5 km/s. To_LSR_Kinematic add -21 km/s
%              B 0312-770                add  -8.3 km/s. To_LSR_Kinematic add -18.5 km/s
%              ESO 55- 2                add  -6.0 km/s. To_LSR_Kinematic add -20.0 km/s
%              PKS 0355-66              add  -4.0 km/s. To_LSR_Kinematic add -19.5 km/s.

\subsection{H\,{\sc i} 21-cm observations}
\label{hidata}

Previously-existing 21-cm H\,{\sc i} data were taken from two sources. One sightline 
towards B\,0242--7229 was previously observed by Muller et al. (2003b), and combines 
ATCA and Parkes data. This has a velocity channel separation of $\sim$1.6 km\,s$^{-1}$ 
and a spatial resolution of 98 arcsec. Assuming a distance to the Magellanic Bridge 
of $\sim$ 55 kpc (e.g. see Muller et al. 2003b), this corresponds to a linear scale of
$\sim$25 pc at the Bridge distance. The final spectrum has a RMS noise of 1.2 K. 
All seven sightlines also have Parkes-only observations of the Magellanic Bridge 
taken from Br\"uns et al. (2005). Each of these pointings is within 3 arcmin of the 
target QSOs. These data were taken during 1999 and used the multibeam L--band front-end 
on the Parkes radiotelescope with 
frequency switching. These data have a spatial resolution of 14.1 
arcmin (corresponding to $\sim$225 pc at the distance of the Magellanic Bridge), 
and a velocity resolution of 0.8 km\,s$^{-1}$. They were reduced using standard
methods, which included conversion to a flux scale using observations of 
S8 and S9, removal of the baseline by subtracting in most cases a linear baseline 
from the bandpass-corrected data, removal of interference spikes 
and conversion to the LSR. 

\begin{figure}
\includegraphics[]{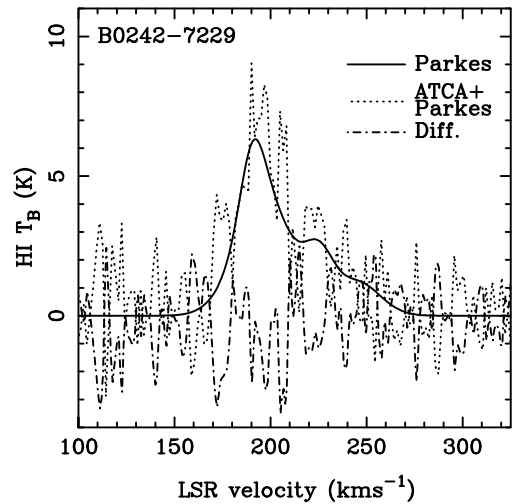}
\caption{Comparison between 21-cm H\,{\sc i} ATCA-plus-Parkes observations 
taken from Muller et al. (2003b), 
and Parkes-only data taken from Br\"uns et al. (2005), towards the QSO B\,0242-7229.}
\label{ATCAvsParkes}
\end{figure}

A comparison between the ATCA-plus-Parkes and Parkes-only observations towards 
B\,0242--7229  is shown in Fig. \ref{ATCAvsParkes}, plus the difference between the 
two datasets. For this particular sightline at least, and at this 
signal-to-noise, the difference between the profiles is negligible. The 
total Bridge H\,{\sc i} column density measured using the two sightlines is 
similar, being 47$\times$10$^{19}$ cm$^{-2}$ for the Parkes-only data, and
52$\times$10$^{19}$cm$^{-2}$ for the ATCA-plus-Parkes spectrum. 

\begin{figure*}
\includegraphics[]{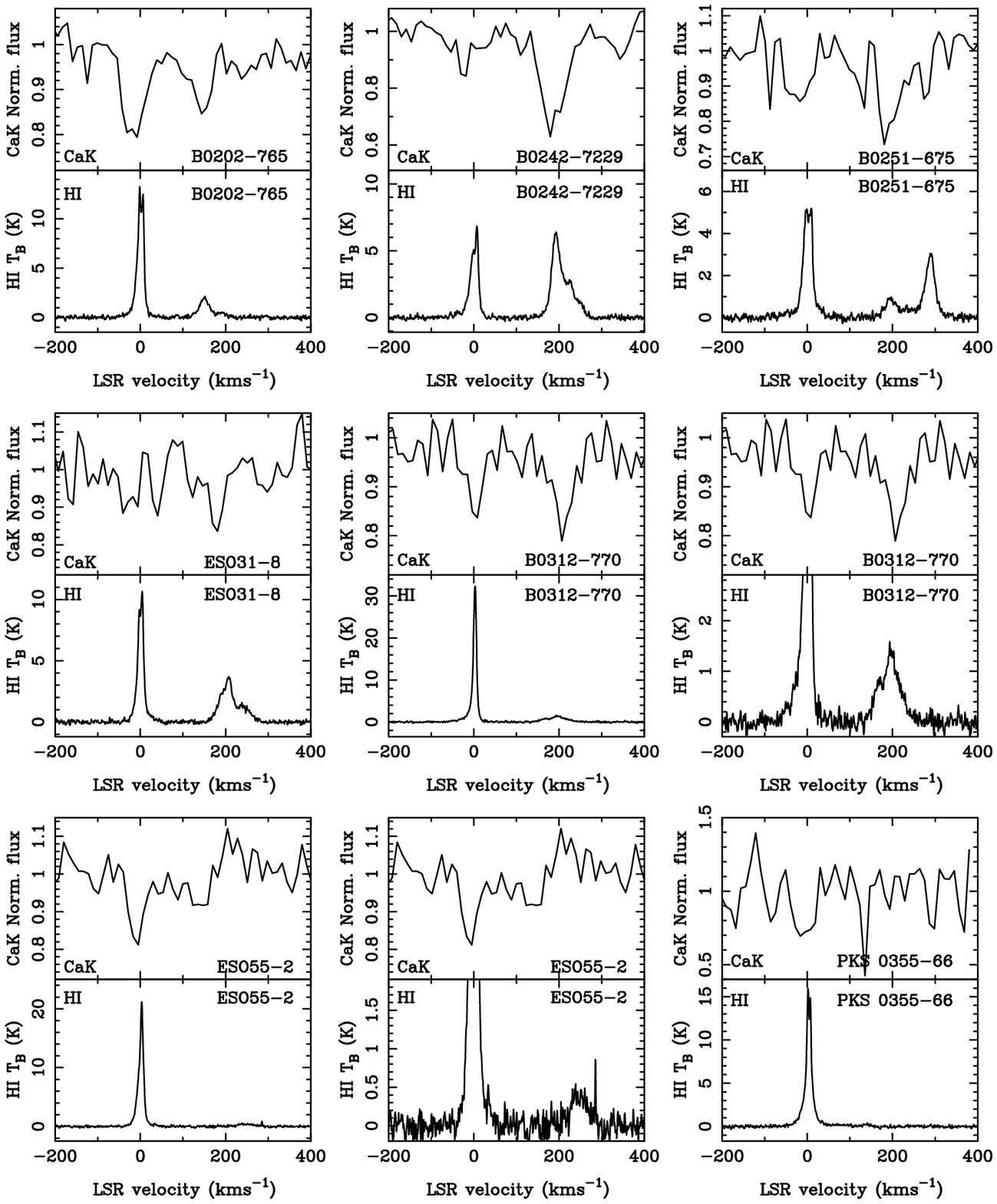}
\caption{Comparison between {\sc emmi} Ca\,{\sc ii} K 
absorption-line and Parkes 21-cm H\,{\sc i} 
emission-line observations of the individual sightlines. 
In the cases of B\,0312--770 and ESO\,55--2, 
two plots are shown in order to highlight differences in H\,{\sc i} strength.}
\label{spectra}
\end{figure*}

%\clearpage
%\newpage
\section{Data analysis and results}
\label{analysis}

\subsection{Ca\,{\sc ii} K and H\,{\sc i} spectra}
\label{CaKHIanalysis}

Figure \ref{spectra} shows the Ca\,{\sc ii} K spectra of the seven lines of 
sight, plotted in the kinematical LSR. Also 
shown in the 
figure are the H\,{\sc i} emission-line spectra 
described in Sect \ref{hidata}. 
Immediately apparent is that the strength of Ca\,{\sc ii} K in the Bridge is 
much greater at a 
constant $N$(H\,{\sc i}) than in low-velocity Galactic gas. The difference was 
quantified by using the {\sc elf} 
suite of programmes within {\sc dipso} in order to measure 
the equivalent widths, central 
velocities and velocity widths of Ca\,{\sc ii} K and H\,{\sc i} 
components. Equivalent width results were compared for a few objects to the quick-look reduction 
performed during the observing run, and the results were found to be the same within 20 per cent. 
The results from the current reduction of the data described in Sect. \ref{observations} are 
those used in the results.

For Ca\,{\sc ii} 
K the column density was estimated from the equivalent width ($EW$) via:

\begin{equation}
N({\rm cm^{-2}})=1.13 \times 10^{17} \times \frac{EW ({\rm m\AA})}{f \times \lambda^{2}
({\rm \AA)}},
\label{coldensVerg}
\end{equation}

where $\lambda$ is the rest wavelength of the line and $f$ is the oscillator strength 
(e.g. 
Spitzer 1978). The Ca\,{\sc ii} K oscillator strength of 0.627 and 
rest wavelength of 3933.661\AA \, were taken 
from Morton (2003, 2004). 
Equation \ref{coldensVerg} is only valid in the optically-thin approximation, and
hence the derived 
column densities are lower limits. 
Due to the poor velocity resolution, unresolved Ca\,{\sc ii} K 
absorption features will almost certainly be present.

\begin{figure*}
\includegraphics[]{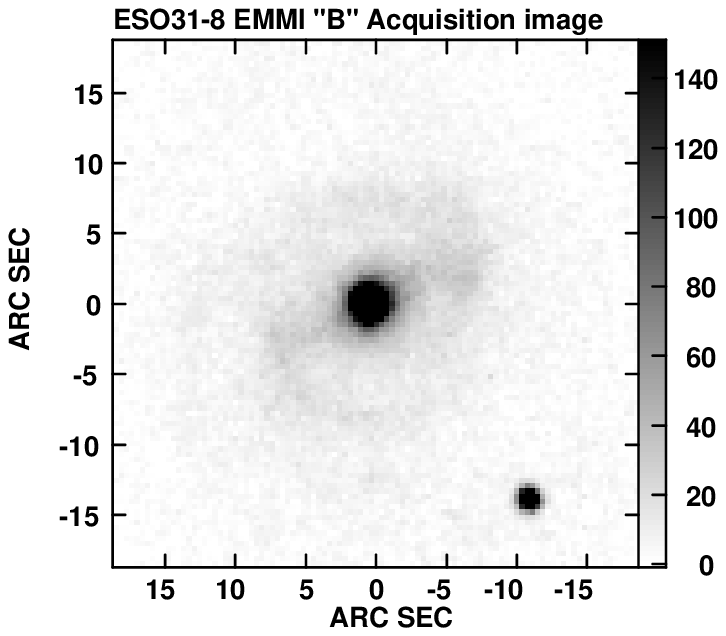}
\includegraphics[]{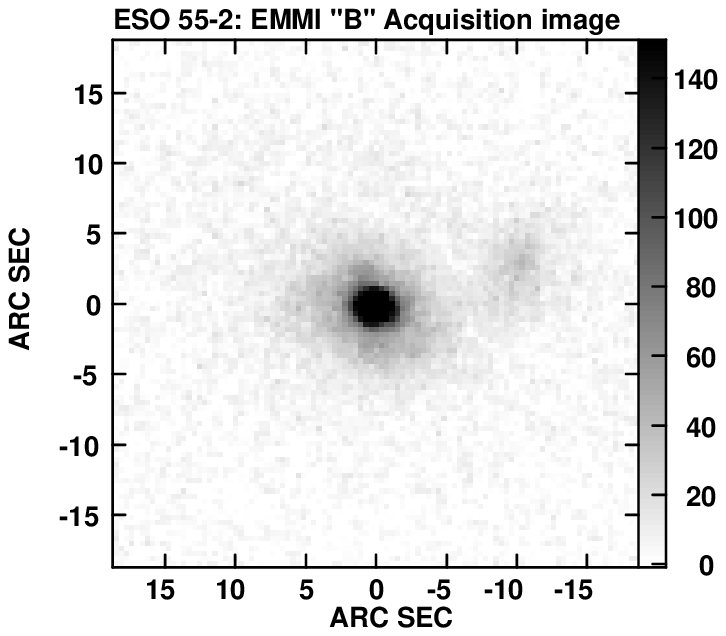}
\caption{{\sc emmi} $B$-band acquisition images of two extended objects in the sample. North is up and
East is to the left. The peak level for the ESO\,31--8 image in the Seyfert nucleus is 5200 counts above sky 
and for ESO\,55-2 is 820 counts above sky.}
\label{BImages}
\end{figure*}

\begin{figure*}
\includegraphics[]{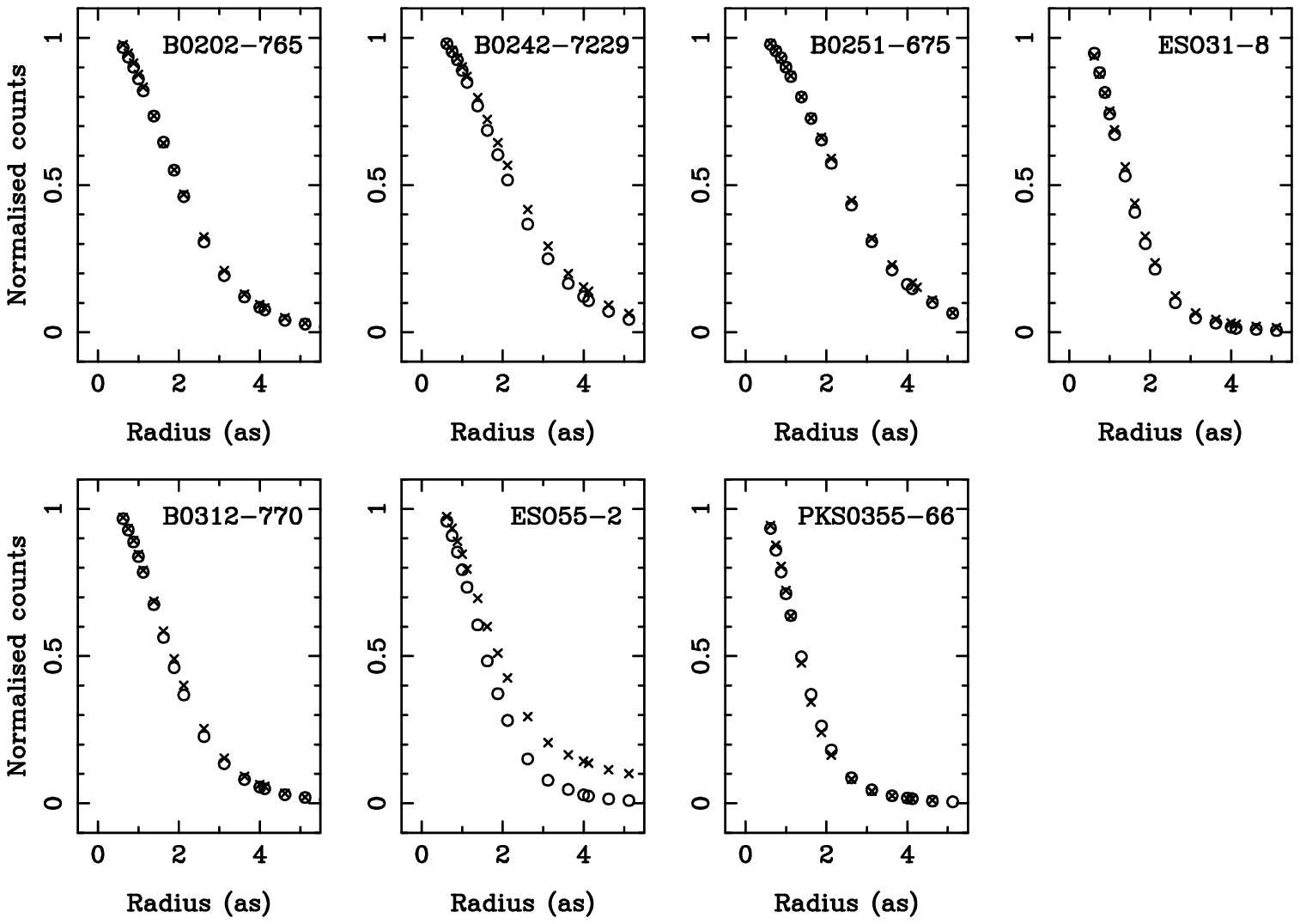}
\caption{{\sc emmi} $B$-band acquisition images: profile fits to the sample objects. 
Open circles refer to fits to nearby stars and crosses to fits to the QSOs. When the points 
lie on top of each other this implies that the QSOs are unresolved by the current observations.}
\label{proffits}
\end{figure*}

For the H\,{\sc i} data the column density 
$N$(H\,{\sc i}) was estimated thus;

\begin{equation}
N({\rm HI}) ({\rm cm^{-2}})=1.823\times 10^{18} \times \int_{v1}^{v2}{T_{\rm B} dv}
\label{coldensHI}
\end{equation}

%
% Two-component ELF-fit to 024311+721630 ATCA-plus-Parkes data
%
% Line Profile    Centre   Width       Peak Flux   Line Flux
%   1     1       204.150   27.459     5.631E+00   1.646E+02
%       +/-         0.000    1.144     0.000E+00         NAN
%   2     1       230.835   44.184     2.350E+00   1.105E+02
%       +/-         0.000    0.000     0.000E+00         NAN
%

where $T_{\rm B}$ is the brightness temperature in K integrated over 
the velocity range from $v1$ to $v2$ in km\,s$^{-1}$. 
The results from {\sc elf} and the derived column densities for the individual 
components 
are shown in Table \ref{resultstable1}, and the corresponding total column densities
in Table \ref{resultstable2}. 

For the sightlines with non-detections, we have estimated the limiting Ca\,{\sc ii} K 
equivalent width from the instrumental FWHM resolution, $\Delta$$\lambda$$_{\rm inst}$=0.66\AA, and the
S/N ratio of the spectra. We can calculate the smallest (3$\sigma$) equivalent width,
$EW_{\rm lim}$, of an unresolved feature that we would expect to see by the
relationship:

$EW_{\rm
lim}$~=3$\times$~$\Delta$$\lambda$$_{\rm inst}$~(S/N)$^{-1}$.

\noindent
These estimated limiting equivalent widths are somewhat high and are not discussed further in the paper.

Table \ref{resultstable2} also gives the ionic abundance relative to 
H\,{\sc i}, $A$=log($N$(Ca\,K)/$N$(H\,{\sc i})). 
It is important to stress that the resolution of the
optical and radio observations are totally different, 
with the Parkes 
\begin{table*}
\begin{center}
\small
\caption{Comparison between H\,{\sc i} emission-line and Ca\,{\sc ii} K absorption line data. }
\label{resultstable1}
\begin{tabular}{lrrrrrrrrrrrrrr}
\hline
    QSO      &   $v_{\rm HI}$ &  FWHM(H\,{\sc i}) &T$_{B}^{\rm max}$(H\,{\sc i})&  H\,{\sc i} flux  & $v_{\rm CaK}$   & EW(CaK)           & S/N ratio \\
 sightline   & (km\,s$^{-1}$) & (km\,s$^{-1}$)&   (K)           &  (K km\,s$^{-1}$) & (km\,s$^{-1}$)  & (m\AA)            & (CaK) \\
\hline
             &                &               &                 &                   &                 &                   &       \\
 B\,0202--765  & --4.9$\pm$0.6  &  30.1$\pm$1.2 &   2.27$\pm$0.26 &  72.9$\pm$6.1     &  --14.7$\pm$3.1 & 174.5$\pm$18.0  &  31   \\
    ``         &   0.4$\pm$0.1  &  14.8$\pm$0.3 &   8.68$\pm$0.25 & 137.3$\pm$6.5     &         --      &       --        &       \\
    ``         & --1.6$\pm$0.0  &   2.7$\pm$0.1 &   2.91$\pm$0.11 &   8.4$\pm$0.5     &         --      &       --        &       \\
    ``         &   6.7$\pm$0.0  &   4.1$\pm$0.1 &   5.34$\pm$0.12 &  23.6$\pm$1.0     &         --      &       --        &       \\
    ``         & 150.0$\pm$0.2  &  28.0$\pm$0.6 &   1.86$\pm$0.03 &  55.8$\pm$1.0     &  +139.2$\pm$3.5 & 122.1$\pm$15.4  &       \\
    ``         & 190.2$\pm$0.8  &  22.4$\pm$2.3 &   0.45$\pm$0.03 &  10.8$\pm$0.9     &         --      &       --        &       \\
               &                &               &                 &                   &                 &                 &       \\
 B\,0242--7229 &--35.6$\pm$0.0  &  22.7$\pm$0.0 &   0.39$\pm$0.02 &   9.4$\pm$0.6     &  --24.1$\pm$2.9 &  50.1$\pm$16.4  &  21   \\
    ``         & --0.3$\pm$0.0  &  22.7$\pm$0.0 &   4.75$\pm$0.02 & 115.2$\pm$0.7     &         --      &       --        &       \\
    ``         &   7.6$\pm$0.0  &   5.6$\pm$0.0 &   3.40$\pm$0.05 &  20.6$\pm$0.3     &         --      &       --        &       \\
    ``         & 190.5$\pm$0.2  &  14.6$\pm$0.7 &   2.40$\pm$0.20 &  37.5$\pm$4.7     &  +185.7$\pm$2.5 & 282.8$\pm$24.9  &       \\
    ``         & 196.6$\pm$0.4  &  33.6$\pm$0.9 &   4.19$\pm$0.19 & 150.3$\pm$3.7     &         --      &       --        &       \\
    ``         & 225.0$\pm$0.0  &  18.0$\pm$0.6 &   1.81$\pm$0.08 &  34.8$\pm$2.1     &         --      &       --        &       \\
    ``         & 245.0$\pm$0.0  &  27.9$\pm$0.9 &   1.20$\pm$0.03 &  35.7$\pm$1.1     &         --      &       --        &       \\
               &                &               &                 &                   &                 &                 &       \\
 B\,0251--675  &   2.1$\pm$0.7  &  56.0$\pm$3.8 &   0.86$\pm$0.10 &  51.8$\pm$3.4     &  --21.6$\pm$7.4 & 120.1$\pm$32.3  &  19   \\
    ``         &   9.7$\pm$0.1  &   7.0$\pm$0.4 &   2.83$\pm$0.20 &  21.2$\pm$2.8     &         --      &       --        &       \\
    ``         & --1.3$\pm$0.3  &  17.7$\pm$0.9 &   4.24$\pm$0.10 &  80.2$\pm$6.0     &         --      &       --        &       \\
    ``         & 194.4$\pm$0.5  &  26.7$\pm$1.8 &   0.68$\pm$0.03 &  19.6$\pm$1.8     &  +194.5$\pm$9.1 & 268.8$\pm$52.5  &       \\
    ``         & 250.9$\pm$4.5  & 110.9$\pm$6.3 &   0.39$\pm$0.01 &  46.2$\pm$2.7     &         --      &       --        &       \\
    ``         & 288.7$\pm$0.1  &  24.4$\pm$0.4 &   2.64$\pm$0.03 &  68.8$\pm$1.4     &         --      &       --        &       \\
               &                &               &                 &                   &                 &                 &       \\
 B\,0312--770  &   2.8$\pm$0.0  &   7.8$\pm$0.0 &  29.84$\pm$0.25 & 249.8$\pm$3.5     &     2.1$\pm$4.1 &  87.1$\pm$18.8  &  30   \\ 
    ``         & --4.8$\pm$0.4  &   9.0$\pm$0.8 &   2.94$\pm$0.19 &  28.5$\pm$4.3     &         --      &       --        &       \\
    ``         & --4.6$\pm$0.3  &  37.8$\pm$1.3 &   2.13$\pm$0.12 &  86.2$\pm$2.4     &         --      &       --        &       \\
    ``         &+168.3$\pm$1.0  &  28.4$\pm$8.0 &   0.15$\pm$0.07 &   4.6$\pm$1.5     &  +208.5$\pm$4.2 & 141.8$\pm$23.2  &       \\
    ``         &+168.3$\pm$2.0  &  10.0$\pm$1.0 &   0.13$\pm$0.08 &   1.4$\pm$0.5     &         --      &       -         &       \\
    ``         &+195.8$\pm$1.0  &  64.0$\pm$2.1 &   0.87$\pm$0.05 &  59.3$\pm$2.7     &         --      &       --        &       \\
    ``         &+197.2$\pm$0.6  &  17.4$\pm$2.0 &   0.55$\pm$0.06 &  10.1$\pm$1.9     &         --      &       --        &       \\
               &                &               &                 &                   &                 &                 &       \\
 ESO\,31--8    & --2.5$\pm$0.0  &   5.9$\pm$0.2 &   4.47$\pm$0.15 &  28.5$\pm$1.8     &         --      & $<$200          &  10   \\
    ``         & --0.7$\pm$0.1  &  22.5$\pm$0.4 &   4.45$\pm$0.16 & 106.8$\pm$2.5     &         --      &     "           &       \\
    ``         &   4.4$\pm$0.0  &   6.1$\pm$0.1 &   6.59$\pm$0.12 &  42.9$\pm$1.5     &         --      &     "           &       \\
    ``         &  28.6$\pm$1.3  &  21.2$\pm$3.1 &   0.32$\pm$0.03 &   7.3$\pm$1.0     &         --      &     "           &       \\
    ``         & 198.0$\pm$1.0  &  35.7$\pm$1.1 &   2.52$\pm$0.08 &  96.2$\pm$5.9     &  +179.5$\pm$5.0 &  75.0$\pm$40.0  &       \\
    ``         & 208.8$\pm$0.1  &  11.8$\pm$0.8 &   1.46$\pm$0.11 &  18.4$\pm$2.4     &         --      &       --        &       \\
    ``         & 241.1$\pm$1.4  &  40.3$\pm$2.2 &   1.21$\pm$0.03 &  51.9$\pm$3.8     &         --      &       --        &       \\
               &                &               &                 &                   &                 &                 &       \\
 ESO\,55--2    & --0.8$\pm$0.4  &  14.0$\pm$0.7 &   7.04$\pm$0.64 & 105.2$\pm$9.9     &   --8.8$\pm$3.6 &  91.2$\pm$20.1  &  19   \\
    ``         & --1.2$\pm$0.3  &  29.6$\pm$1.7 &   2.49$\pm$0.43 &  78.8$\pm$9.3     &         --      &       --        &       \\
    ``         &   2.9$\pm$0.0  &   3.4$\pm$0.1 &   4.39$\pm$0.27 &  16.2$\pm$1.6     &         --      &       --        &       \\
    ``         &   4.3$\pm$0.1  &   8.2$\pm$0.2 &   9.44$\pm$0.68 &  82.4$\pm$8.0     &         --      &       --        &       \\
    ``         &  35.3$\pm$0.6  &   8.8$\pm$2.0 &   0.36$\pm$0.05 &   3.4$\pm$0.5     &         --      &       --        &       \\
    ``         & 246.7$\pm$1.1  &  45.4$\pm$2.8 &   0.39$\pm$0.02 &  19.1$\pm$1.0     &         --      & $<$100          &       \\
               &                &               &                 &                   &                 &                 &       \\
 PKS\,0355--66 &--16.0$\pm$2.1  &  21.5$\pm$3.2 &   0.80$\pm$0.10 &  18.5$\pm$4.7     &         --      & $<$330          &   6   \\
    ``         &   3.9$\pm$0.2  &  21.2$\pm$0.4 &   6.78$\pm$0.17 & 153.2$\pm$4.9     &         --      &     "           &       \\
    ``         &   1.6$\pm$0.0  &   5.0$\pm$0.0 &   9.00$\pm$0.13 &  48.6$\pm$1.2     &         --      &     "           &       \\ 
    ``         &   7.3$\pm$0.0  &   3.5$\pm$0.0 &   7.78$\pm$0.12 &  29.8$\pm$0.8     &         --      &     "           &       \\
    ``         &  25.9$\pm$4.2  & 125.9$\pm$8.6 &   0.31$\pm$0.02 &  42.0$\pm$2.4     &         --      &     "           &       \\
\hline
\end{tabular}
\end{center}
\normalsize
\end{table*}
%
%
%
%
% Fudge factor is 1.1653e10 times ew in mA for CaK
%
21-cm H\,{\sc i} data 
sampling a much larger area than the Ca\,{\sc ii} K 
observations, so values of $A$ will 
clearly only be indicative, even (incorrectly) 
assuming that the Ca\.{\sc ii} K spectra have no 
hidden substructure. With these caveats, the values 
of $A$ for low-velocity gas range from --8.3 to --8.8 
dex with a range in H\,{\sc i} column density 
from 3$-$6$\times$10$^{20}$ cm$^{-2}$. For Magellanic 
Bridge gas, the values of $A$ are $\sim$ 0.5 dex higher, 
ranging from $\sim$ --7.8 to --8.2 dex, 
with $N$(H\,{\sc i})=1--5$\times$10$^{20}$ cm$^{-2}$. 

\subsection{Optical images --- profile fits to observed objects}

Previously, researchers have used the extended regions in the
cores of some globular clusters as a background light source to 
probe the structure of low and intermediate velocity interstellar
clouds (IVCs) on arcsecond scales, such as the IVC towards M\,15 
(Meyer \& Lauroesch 1999, Smoker et al. 2002). In order to determine
whether in future we could use objects 
in the current sample to similarly probe the gas in the Magellanic Bridge, 
the $B$-band acquisition images of the sample were inspected for signs of 
extended emission. 
If the objects had turned out to be extended, then follow-up 
observations for example using {\sc flames} on the {\sc vlt} would be 
appropriate. However, only in the cases of ESO\,31--8 and ESO\,55--2 
was there anything other than an unresolved nuclear component apparent in the images 
shown in Fig. \ref{BImages}. 
This was confirmed by using the Extended Surface Photometry ({\sc esp}) 
package (Gray et al. 2000) to fit ellipses to the main target and a 
nearby comparison star. The results are shown in Fig. \ref{proffits} which plots the 
extracted counts as a function of increasing radius, using a centre fixed 
in (x,y). Although ESO\,31--8 {\em does} have a large extended component out 
to $\sim$ 5 arcsec radius, it is a factor $\approx$ 100 weaker than the nuclear peak, 
and would not be adequately bright for our purposes to probe the Bridge gas 
on arcsecond scales. ESO\,55--2 may be a good candidate for follow-up projects to probe the 
extended absorption properties of the Bridge ISM given that its core 
extends to $\sim$ 1.8 arcseconds with a brightness of 1/10th of the peak. 

\subsection{A note on ESO\,55--2}

\begin{figure}
\includegraphics[]{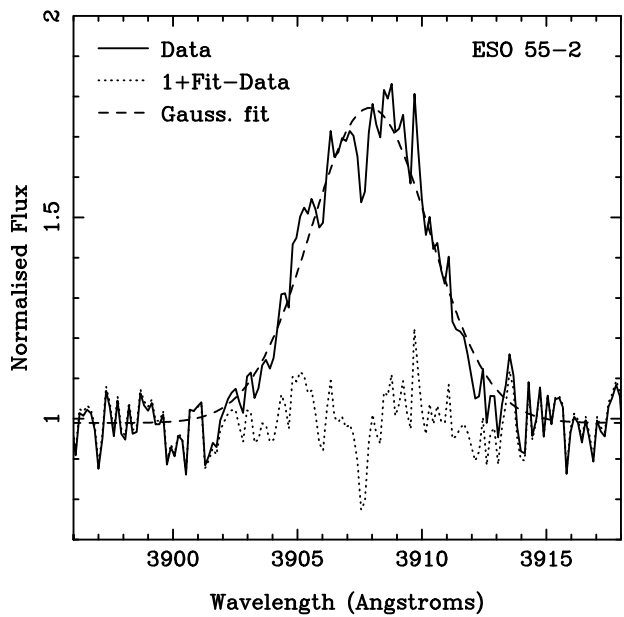}
\caption{ESO\,55--2; normalised {\sc emmi} spectrum in the region of redshifted [O\,{\sc ii}]
($\lambda_{\rm air}$=3727.319\AA). One plus the residual of the Gaussian-fit minus the extracted
spectrum is shown as a dotted line.}
\label{eso55emission}
\end{figure}

ESO\,55--2 is a Seyfert 2 galaxy catalogued by Arp \& Madore (1987). 
It is one of only two objects in the current sample that are well-resolved by the current observations. 
In addition, ESO\,55--2 is the only object for which we detect a clear emission line 
shown in Fig. \ref{eso55emission}.
This is a broad feature and lies 
at a (uncorrected) wavelength of 3907.91$\pm$0.06\AA \, 
with a FWHM width of 5.65$\pm$0.14\AA 
\, (equivalent to $\sim$410 km\,s$^{-1}$ in the restframe), and with a line 
flux of 4.7$\pm$0.2\AA. 
It is likely to be $[$O\,{\sc ii}$]$ at $\lambda_{\rm air}$=3727.319\AA, redshifted to the 
quasar heliocentric velocity of 14538$\pm$74 km\,s$^{-1}$ (Sekiguchi \& Wolsentcroft 1993).

\section{Discussion}
\label{discussion}

There are limitations on the information concerning the physical conditions 
of the gas in the Bridge from the Ca II K and H\,{\sc i} spectra alone, especially
where they have such different spatial resolutions. The main result is that 
in the 
three sightlines with reasonably well-defined values of the 
Magellanic Bridge Ca\,{\sc ii} K absorption, the values of $A$ are some 
$\sim$ 0.5 dex higher than in Galactic gas. We recall that Wakker \& Mathis (2000) (although see Smoker et al. 2003) 
related Ca\,{\sc ii} K to H\,{\sc i} in the following way;

\begin{equation}
{\rm log} A({\rm Ca\,II}) = -0.78 \times ({\rm log}(N({\rm H\,I})) - 19.5) - 7.78 
\end{equation}

\begin{table*}
\begin{center}
\small
\caption{Comparison of H\,{\sc i} emission-line and Ca\,{\sc ii} K absorption line data - total column 
densities. ``Bridge'' refers to Magellanic Bridge gas and ``LV'' to low-velocity gas. As noted in the 
text, there are almost certainly unresolved Ca\,{\sc ii} components, these values are lower limits. 
The Ca\,{\sc ii} K detections towards B\,0251--675 and ESO\,31--8 are tentative.}
\label{resultstable2}
\begin{tabular}{lrrrrrrr}
\hline
   QSO        & Region  & H\,{\sc i}(Tot)  & CaK(Tot)       & $N$(H\,{\sc i})             & $N$(CaK)                     & $A$(CaK) \\
              &         & K km\,s$^{-1}$   & m\AA           & $\times$10$^{19}$ cm$^{-2}$ & $\times$10$^{11}$ cm$^{-2}$  &          \\
\hline
              &         &                  &                &              &               &         \\
B\,0202--765  &  LV     & 242.2$\pm$10.2   & 174.5$\pm$18.0 & 44.2$\pm$1.9 & 20.3$\pm$2.1  & --8.33$\pm$0.06  \\
    ``        & Bridge  &  66.6$\pm$1.9    & 122.1$\pm$15.4 & 12.1$\pm$0.4 & 14.2$\pm$1.8  & --7.93$\pm$0.07  \\
              &         &                  &                &              &               &                  \\
B\,0242--7229 &  LV     & 145.2$\pm$3.1    &  50.1$\pm$16.4 & 26.5$\pm$0.6 &  5.8$\pm$1.9  & --8.65$\pm$0.12  \\
    ``        & Bridge  & 258.3$\pm$8.2    & 282.8$\pm$24.9 & 47.1$\pm$1.5 & 33.0$\pm$2.9  & --8.15$\pm$0.05  \\
              &         &                  &                &              &               &                  \\
B\,0251--675  &  LV     & 153.2$\pm$8.0    & 120.1$\pm$32.3 & 27.9$\pm$1.5 & 14.0$\pm$3.8  & --8.30$\pm$0.13  \\
    ``        & Bridge  & 134.6$\pm$4.5    & 268.8$\pm$52.5 & 24.5$\pm$0.8 & 31.3$\pm$6.1  & --7.89$\pm$0.09  \\ 
              &         &                  &                &              &               &                  \\
B\,0312--770  &  LV     & 364.5$\pm$9.4    &  87.1$\pm$18.8 & 66.4$\pm$1.7 & 10.1$\pm$2.2  & --8.81$\pm$0.09  \\
    ``        & Bridge  &  75.4$\pm$4.0    & 208.5$\pm$23.2 & 13.7$\pm$0.7 & 24.3$\pm$2.7  & --7.75$\pm$0.07  \\
              &         &                  &                &              &               &                  \\
ESO\,31--8    &  LV     & 185.5$\pm$16.8   &    --          & 33.8$\pm$3.1 &      $<$23.0  & $<$--8.16        \\
    ``        & Bridge  & 166.5$\pm$8.3    &  75.0$\pm$40.0 & 30.4$\pm$1.6 &  8.7$\pm$4.6  & --8.58$\pm$0.25  \\
              &         &                  &                &              &               &                  \\
ESO\,55--2    &  LV     & 286.0$\pm$16.2   &  91.2$\pm$20.1 & 52.1$\pm$3.0 & 10.6$\pm$2.4  & --8.68$\pm$0.10  \\
    ``        & Bridge  &  19.1$\pm$1.1    &    --          &  3.4$\pm$0.2 &      $<$12.0  & $<$--7.45        \\
              &         &                  &                &              &               &                  \\
PKS\,0355--66 &  LV     & 292.1$\pm$9.4    &    --          & 53.2$\pm$1.8 &      $<$38.4  & $<$--8.13        \\
    ``        & Bridge  &   --             &    --          &   --         &         "     &    --            \\
\hline
\end{tabular}
\end{center}
\normalsize
\end{table*}

\begin{figure}
\includegraphics[]{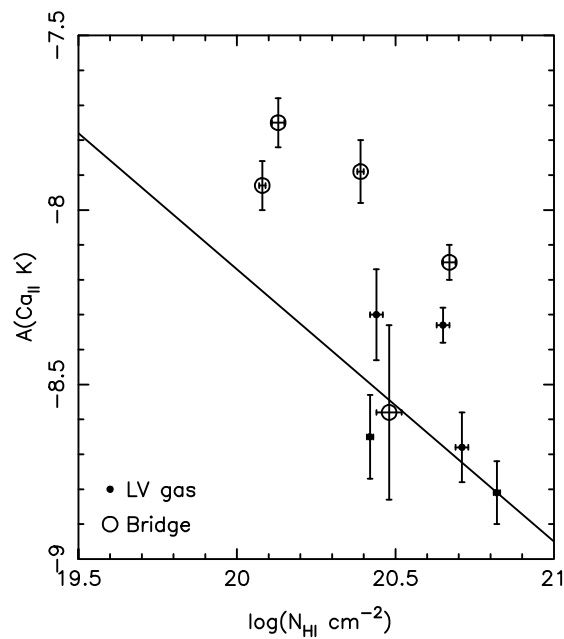}
\caption{Comparison between $A$(Ca\,{\sc ii}) and log($N_{\rm HI}$) for the low-velocity and
Bridge components in the current dataset. The solid line refers to the best fitting value
taken from Wakker \& Mathis (2000). We note that the Ca\,{\sc ii} abundances are likely
lower limits due to the fact that unresolved features will almost certainly be present in
the {\sc emmi} data.}
\label{A_vs_NH}
\end{figure}

Hence for typical Galactic 
gas which has $N$(H\,{\sc i}) of $\sim$ 5$\times$10$^{20}$ cm$^{-2}$, 
we would expect $A$(Ca\,{\sc ii}) $\sim$ --8.7, which is as observed in 
the current 
sightlines. {\em If} the physical properties of Bridge gas were the same as for Galactic
gas, for the Bridge 
sightlines which have $N$(H\,{\sc i}) of $\sim$ 2$\times$10$^{20}$ cm$^{-2}$, 
we would expect $A$(Ca\,{\sc ii}) 
$\sim$ --8.4, which is a smaller by $\sim$ 0.5 dex than is observed in the current Bridge 
dataset. The difference could be caused by a number of conflicting factors including;  
(a) Ionisation issues. If some of the hydrogen is in ionised form in the Bridge 
this would cause $N$(H\,{\sc i}) to be lower in the Bridge and hence $A$ to be higher. 
Observations in H$\alpha$ would be able to determine the 
level of ionisation of the hydrogen in the Bridge and Galaxy components on the current
sightlines. 
(b) the presence of less dust in the Magellanic Bridge than in the Milky Way 
onto which Ca is bound, which causes less depletion and hence would increase the observed 
value of $N$(Ca\,{\sc ii}) and $A$, and (c) a smaller intrinsic metallicity in the 
Magellanic Bridge system when compared with the Milky Way which would decrease 
both $N$(Ca\,{\sc ii}) and $A$. Previous work on 
a Magellanic Bridge B-type star by Rolleston et al. (1999) and interstellar observations 
of the Bridge by Lehner et al. (2001) support a metallicity of $\sim -1.1$ dex compared to 
the Galaxy. Fig. \ref{A_vs_NH} displays $A$(Ca\,{\sc ii} K) against $N$(H\,{\sc i}) for the 
low-velocity and Bridge gas respectively. Also plotted on the diagram is the best-fitting 
line taken from Wakker \& Mathis (2000). The current low velocity points scatter about
this best-fitting line. As noted above, the Magellanic Bridge points are offset by about 
0.5 dex, although to first order the slope of the line (admittedly heavily biased by the 
observation towards B\,0242-7229) seems to be the same as for the low-velocity gas. 

Concluding, the observed value of $A$ is affected by many variables, and the current data
alone are not sufficient to unscramble the different effects. 

One sightline, B\,0251-675, is particularly interesting and merits further study. 
This position lies approximately mid-way between the SMC and LMC and has two clear 
velocity components visible in H\,{\sc i} at $\sim$ +195 and +289 km\,s$^{-1}$, 
joined by a Bridge of gas with velocity $\approx$ +249 km\,s$^{-1}$. Although the line 
flux of the +289 km\,s$^{-1}$ 
component is some 3.7 times stronger than that at 195 km\,s$^{-1}$, 
Fig. \ref{spectra} shows obvious absorption only in the lower velocity component. 
The H\,{\sc i} component at +289 km\,s$^{-1}$ has a line flux of 69 K km\,s$^{-1}$ or 
1.3$\times$10$^{20}$ cm$^{-2}$. This could imply that at this position, the Bridge is 
composed of gas which has different origins, or currently has different ionisation or dust properties. 

It appears that towards the B\,0251--675 sightline, we could be probing two
different ``arms'' of the SMC, Bridge or Magellanic Stream. The two components are separated by a few kpc
and are both visible in H\,{\sc  i}.  The lower velocity arm is obviously
turbulent and morphologically complex (Muller et al. 2004), and numerical
simulations by Gardiner et al. (1994) suggest this component forms the actual
``Bridge'' link between the SMC and LMC. The higher velocity component is an
almost radially extending arm, and has a much smoother appearance in H\,{\sc
  i}.

The continued tidal interactions between the Magellanic Clouds is the most
likely mechanism for the mixing of the H\,{\sc i} in the Magellanic
Bridge. This feature is also apparently more chemically enriched, relative to
the higher velocity component, as suggested in the numerical simulations by
Gardiner et al. (1994) and by our EMMI observations.

Higher S/N ratio Ca\,{\sc ii} data combined with high angular-resolution H\,{\sc i} observations 
would be needed to confirm the current finding because there remains the possibility that
there {\em is} H\,{\sc i} gas towards the B\,0251--675 that may only be detected using a pencil 
beam where at present it is diluted by the large Parkes resolution. Higher resolution observations 
are thus necessary to more convincingly associate the absorption feature with a line of sight H\,{\sc i} 
component.

\section{Conclusions and future work}
\label{conclusions}

We have presented intermediate spectral resolution Ca\,{\sc ii} K and 
high spectral resolution H\,{\sc i} observations towards 
7 QSO sightlines in the Magellanic Bridge. 
Clearly the current observations have provided only a 
first-look towards QSO Bridge sightlines, as the signal-to-noise ratios 
in the Ca\,{\sc ii} K spectra 
are at the limit for what can be usefully derived and the optical and 
H\,{\sc i} data are not well matched in resolution. In the three objects with 
good S/N ratio, we find that the ionic abundance of Ca\,{\sc ii} K relative to H\,{\sc i} 
is $\sim$ 0.5 dex higher than for local gas, assuming that the H\,{\sc i} 
column density on arcsecond scales matches that at the Parkes resolution 
of 14 arcminutes and of course in the absence of ionisation corrections 
or differences in dust content. 

Obvious ground-based follow-up observations to the present work 
would be higher-resolution spectroscopy in Ca\,{\sc ii} K and 
the Na\,D lines on a 8-m telescope, for example using {\sc uves} on the {\sc vlt}. 
This would both enable a higher S/N to be attained, and also to easily distinguish 
different Bridge components, especially towards B\,0251--675, determining whether the 
apparent difference in the abundance observed in the two velocity components is in 
fact real. Finally, with very good seeing, the {\sc argus} mode of {\sc flames} could be 
used to probe the arcsecond scale structure of Bridge gas, using the Sy2 
ESO\,55--2 as a background source. 

\begin{acknowledgements}
We acknowledge the help of Chip Kobulnicky in preparing the original 
proposal. HMAT and IH acknowledge financial support from the 
Department for Education and Learning for Northern Ireland, 
while JKL is supported by the Particle Physics and Astronomy
Research Council of the United Kingdom. FPK is grateful to AWE 
Aldermaston for the award of a William Penney Fellowship. 
This research has made use of the {\sc simbad} database, operated 
at CDS, Strasbourg, France. We would like to thank Emanuela Pompei 
and Cedric Foellmi for support during our enjoyable run at La Silla. We 
thank an anonymous referee for their useful suggestions for improvement to 
the paper. 

\end{acknowledgements}

{}

\end{document}